\documentstyle[12pt]{article}
\begin{document}
%%%%%%%%%%%%%%%%%%%%%%%%%%%%%%%%%%%%%%%%%%%%%%%%%%%
\def\thefootnote{\fnsymbol{footnote}}
\begin{flushright}
KANAZAWA-96-04  \\ 
November, 1996(Revised)
\end{flushright}
%\vspace{ .7cm}
\vspace*{2cm}
\begin{center}
{\LARGE\bf Neutrino Oscillation based on the Mixings with a 
Heavy Right-Handed Neutrino}\\
\vspace{1 cm}
{\Large  Daijiro Suematsu}
\footnote[3]{e-mail:suematsu@hep.s.kanazawa-u.ac.jp}
\vspace {1cm}\\

{\it Department of Physics, Kanazawa University,\\
        Kanazawa 920-11, Japan}

%{\it and}
%
%{\it Theory Division, CERN, CH-1211 Geneve 23, Switzerland}    
\end{center}
\vspace{2cm}
{\Large\bf Abstract}\\  
%%%%%%%%%%%%%%%%%Abstract%%%%%%%%%%%%%%%%%%%%%%%%%%%
We study a consistent explanation for both deficits of solar neutrino
and atmospheric neutrino due to the neutrino oscillation induced by the mixing 
of light neutrinos with a heavy right-handed neutrino.
We propose such a phenomenological neutrino mass matrix
that realizes this scenario in the simple three generation left- and right-
handed neutrino framework.
Although this model contains only one nonzero mass eigenvalue $\sim 10$~eV
for these light neutrino states at the first approximation level, 
it can be expected to explain consistently both deficiency due to
the appropriate higher order corrections.
A suitable hot dark matter candidate is naturally included in it
as its own feature.

\newpage
\setcounter{footnote}{0}
\def\thefootnote{\arabic{footnote}}
%%%%%%%%%%%%%%%%%Text%%%%%%%%%%%%%%%%%%%%%%%%%%%%%%%
It is one of the most interesting issues to clarify the neutrino mass
problem in the present particle physics.
It has the great influences not only to the consideration of new physics beyond
the standard model but also to the study of astrophysics.
Now we have some clues for this problem. 
The deficiency of solar neutrinos\cite{solar}
and atmospheric neutrinos\cite{atm} have been shown to be explained by 
$\nu_e \rightarrow \nu_x$ and $\nu_\mu \rightarrow \nu_y$ 
oscillation, respectively.
The predicted neutrino masses and mixings from these observations are 
followings.
For solar neutrino problem\cite{mass1},\footnote{
There is another vacuum oscillation solution.
The oscillation parameters for this solution are 
$\Delta m^2\sim (0.5-1.1)\times 10^{-10}{\rm eV}^2,~ 
\sin^22\theta\sim 0.8-1.0.$ However, we do not consider this solution in this
paper.}\\
(i)~MSW small mixing solution:
\begin{eqnarray}
&&\Delta m_{x e}\equiv (m^2_{\nu_x}-m^2_{\nu_e})\sim (0.3-1.2)\times 
10^{-5}{\rm eV}^2, \nonumber \\
&&\sin^22\theta\sim (0.4-1.5)\times 10^{-2},
\end{eqnarray}
(ii)~MSW large mixing solution:
\begin{eqnarray}
&&\Delta m_{x e}\equiv (m^2_{\nu_x}-m^2_{\nu_e})\sim (0.3-3)\times 
10^{-5}{\rm eV}^2, \nonumber \\
&&\sin^22\theta\sim 0.6-0.9,
\end{eqnarray}
and for atmospheric neutrino problem\cite{mass2},
\begin{eqnarray}
&&\Delta m_{y\mu}\equiv |m^2_{\nu_y}-m^2_{\nu_\mu}|\sim (0.5-50)\times
10^{-2}{\rm eV}^2, \nonumber \\
&&\sin^22\theta ~{^>_\sim}~ 0.5.
\end{eqnarray}
Usually it is considered that $\nu_x$ is $\nu_\mu$ and $\nu_y$ is $\nu_\tau$.
It is a very interesting subject to realize these parameters in the suitable
particle physics model.
There are various studies of the neutrino mass matrix which can explain 
these values consistently\cite{bs,matrix}.

On the other hand, it is well known now that the most amount of mass in the 
universe must be stored in the nonbaryonic dark matter.
There have been proposed two possible candidates of dark matter, 
that is, the hot dark matter and the
cold dark matter. However, the analyses based on the data on the structure
of the universe which are now available on a wide range of distance scales 
have shown that these dark matter models fail to explain the structures on
small or large distance scales, respectively. 
It was suggested recently that a cold + hot dark matter
model agrees well with astrophysical observations if there is one neutrino
species with $\sim 5$~eV mass\cite{dark,phkc}. 
From this cosmological point of view, it is a well motivated subject to
introduce $\sim$10~eV neutrino as a candidate of dark matter into the
particle physics model in the natural way.
In fact, there are many trials of the model building to predict 
these neutrino masses\cite{model1}.
However, it seems not to be so easy to construct models which accomodates 
these features naturally.

Usually either the seesaw mechanism\cite{seesaw} or the loop effects 
are used to realize the sufficiently small neutrino masses.
In the ordinary seesaw mechanism, three right-handed neutrinos get the large
Majorana masses and the hierarchy $m_{\nu_\tau}\gg m_{\nu_\mu}\gg m_{\nu_e}$
appears. However, in such a scheme it is rather difficult to
explain the above mentioned observations, simultaneously\cite{bs,matrix}.
On the other hand, if we only use the loop effects, 
the model often becomes complex and it is necessary to
introduce various exotic fields, that is, extra color triplets,
doubly charged fields and so on\cite{loop}.
Under this situation it will be worthy to examine alternative
possibility for the neutrino mass matrix which may explain the small 
neutrino masses and mixings
appropriate for the above mentioned neutrino problems.
 
In this paper we propose a scenario which can accommodate the above mentioned 
features and also can be embedded into the framework similar to
the standard model  with three generation left- and right-handed
neutrinos.
Our strategy is the following. Firstly we require the realization of
the suitable mixings and the dark matter candidate mass and
after that we introduce the appropriate mass differences without
violating the feature of above mixings. 
In this scenario only one right-handed neutrino $N_R$
has a large Majorana mass
and remaining light neutrinos have small mixings with it.
Although there is only one nonzero mass eigenvalue among these
light neutrinos at this first approximation stage due to the 
seesaw mechanism\cite{seesaw},
the required neutrino oscillations can occur if suitable mass 
perturbations expected from these mixings are introduced.   

To discribe the basic feature of this scenario, we consider the model 
defined by the following effective mass terms in the first approximation, 
\begin{equation}
-{\cal L_{\rm mass}}=\sum_{\alpha=1}^4 m_\alpha\psi_\alpha N_{R}
    +{1 \over 2}MN_{R}N_{R}+h.c.,
\end{equation} 
where $\psi_\alpha$ represents the charge conjugate states of
ordinary left-handed neutrinos 
$\psi_{Li}(i=1 \sim 3)$ and also a right-handed sterile one 
$\psi_R$. These and $N_R$ are the weak 
interaction eigenstates and $m_\alpha \ll M$ is assumed.
This can be straightforwardly embedded into the full three
generation neutrino model without any large changes in the following results.
The rank of a $5\times 5$ mass matrix derived from this ${\cal L}_{\rm 
mass}$ is two. One nonzero eigenvalue is large enough compared with
the other one. Using seesaw mechanism we can resolve
the mixing between a heavy state and the remaining four light states.
Under such a basis the mass matrix of four light states are
written as 
\begin{equation}
M_{\rm light}=M\left( \begin{array}{cccc}
\mu_1^2 & \mu_1\mu_2 & \mu_1\mu_3 & \mu_1\mu_4 \\
\mu_1\mu_2 & \mu_2^2 & \mu_2\mu_3 & \mu_2\mu_4 \\
\mu_1\mu_3 & \mu_2\mu_3 & \mu_3^2 & \mu_3\mu_4 \\
\mu_1\mu_4 & \mu_2\mu_4 & \mu_3\mu_4 & \mu_4^2 \\
\end{array}
\right),
\end{equation} 
where $\mu_\alpha= m_{\alpha}/M (\ll 1)$.
As is easily checked, $M_{\rm light}$ is
diagonalized as $U^{(\nu)}M_{\rm light}U^{(\nu)T}$
by using the matrix
\begin{equation}
U^{(\nu)}=\left(
\begin{array}{cccc}
{\mu_2 \over \xi_1} & -{\mu_1 \over \xi_1} & 0 & 0 \\
{\mu_1\mu_3 \over \xi_1\xi_2} & {\mu_2\mu_3 \over \xi_1\xi_2} &
-{\xi_1 \over \xi_2} & 0 \\
{\mu_1\mu_4 \over \xi_2\xi_3} & {\mu_2\mu_4 \over \xi_2\xi_3} &
{\mu_3\mu_4 \over \xi_2\xi_3} & -{\xi_2 \over \xi_3}\\
{\mu_1 \over \xi_4} & {\mu_2 \over \xi_4} & {\mu_3 \over \xi_4} &
{\mu_4 \over \xi_4} \\
\end{array} \right),
\end{equation}
where $\displaystyle \xi_n^2=\sum_{\alpha=1}^{n+1}\mu_\alpha^2$.
At this stage one of the mass eigenvalues is $M\mu_4^2$ and others are
zero and then the constraints for the squared mass differences (1) $\sim$ (3)
cannot be satisfied.
In order to prepare the suitable mass differences among 
these light neutrinos without violationg the mixing property shown 
above, we will introduce a mass perturbation,
\begin{equation}
M_{\rm per}=
U^{(\nu)T}\left(
\begin{array}{cccc}
0&0&0&0\\
0&M_1&0&0\\
0&0&M_2&0\\
0&0&0&0\\
\end{array} \right)U^{(\nu)}.
\end{equation}
The eigenvalues of the perturbed mass matrix $M_{\rm light}+M_{\rm
per}$ are $0, M_1, M_2$ and $M\mu_4^2$.

Now we consider the oscillation phenomena in these four light states.
The mass eigenstates $\phi_\alpha$ is related to the weak interaction
eigenstates $\psi_\alpha$ by the KM-mixing matrix $V^{(l)}$ for 
leptons as
\begin{equation}
\phi_\alpha =\sum_{\beta=1}^4V^{(l)}_{\alpha\beta}\psi_\beta, 
\end{equation} 
where $V^{(l)}$ can be written as 
$V^{(l)}=U^{(\nu)}U^{(l)\dagger}$ by using the diagonalization 
matrix $U^{(l)}$ of the charged lepton mass matrix.
Here we consider the basis on which both of the charged lepton mass
matrix and the lepton charged current are diagonal.
Using this mixing matrix, we can write down the time evolution equation 
of these states, 
\begin{equation}
i{d \over dt}\psi_\alpha = \sum_{\beta=1}^4 H_{\alpha\beta}\psi_\beta.
\end{equation} 
If we assume as usual that these states are ultra relativistic, 
Hamiltonian $H$ is expressed as
\begin{equation}
H_{\alpha\beta}=\sum_{\gamma=1}^4V^{(l)\dagger}_{\alpha\gamma} 
{\tilde m^2_\gamma \over 2E}
V^{(l)}_{\gamma\beta}+a_\alpha\delta_{\alpha\beta},
\end{equation}
where $\tilde m_\gamma$ is the $\gamma$-th mass eigenvalue of $M_{\rm
light}+M_{\rm per}$.
$E$ is the energy of neutrinos.
The potentials induced effectively through the weak interactions with matter
are introduced as $a_\alpha$, which should be taken as zero for the sterile 
neutrinos.  
In order to explain the neutrino oscillations and the dark matter mass
introduced in the first part, 
we need to identify these four light states with the physical ones
 in our neutrino model.
In the following discussion we will focus our attention on the case,
\begin{center}
$\psi_4 \rightarrow$ a right-handed sterile 
neutrino($\nu_s$),
~~$\psi_1, \psi_2, \psi_3 \rightarrow \nu_{eL}, \nu_{\mu L}, \nu_{\tau L}$.
\end{center}

Under this identification to realize the suitable mixing angles 
for the solar and 
atmospheric neutrino problems and also the mass of the dark matter
candidate simultaneously, it is necessary
to require the following relations for each solution (i) 
and (ii),\footnote{Here we assumed $V^{(l)}=U^{(\nu)}$ .
It should be also noted that $\mu_2\sim\mu_3\ll
\mu_4$ is necessary to satisy the constraints (1)$\sim$(3) for mixings 
because of the form of mixing matrix (6).}
\begin{eqnarray}
{\rm (i)}&&16~{^<_\sim}~ {\mu_2 \over \mu_1}~{^<_\sim}~ 32, 
\qquad 1~{^<_\sim}~{\mu_3 \over \mu_2}~{^<_\sim}~ 2.4,  
\qquad {\mu_4 \over \mu_3} \sim 10^p,\nonumber \\
{\rm (ii)}&& 1.4~{^<_\sim}~{\mu_2 \over \mu_1}~{^<_\sim}~2.1, 
\qquad 1~{^<_\sim}~{\mu_3 \over\mu_2}~{^<_\sim}~ 3, 
\qquad {\mu_4 \over \mu_3}\sim 10^p, 
\end{eqnarray}
and the suitable mass as the dark matter candidate, 
\begin{equation}
M\mu_4^2\sim 10~{\rm eV},
\end{equation}
for the largest mass eigenvalue.\footnote{This value should be 
understood as the roughly required order of maginitude.
More detailed analysis will be presented in the later part.} 
At this stage a positive constant $p$ is a free parameter but it
crucially depends on the BBN constraint\cite{wssok}. 
Both values of $p$ and $M\mu_4^2$ will be changed if the state
identification for the dark matter candidate is altered.
On the other hand, since the introduction of $M_{\rm per}$ 
does not modify the mixing property (6), 
the constraints for the 
mass differences are directly imposed on $M_1$ and $M_2$.
That is, if we assume $M_1 \sim 10^{-3}$~eV and $M_2\sim 10^{-1}$~eV, 
the required values for the squared mass differences in the solar 
and atmospheric neutrino deficits can be built in the mass matrix
$M_{\rm light}+M_{\rm per}$. 
For this mass matrix with the parameter setting (11),
in the $(\psi_2, \psi_3)$ sector the mixing 
angle becomes very large as $\sin^2 2\theta \sim 1$ and the effective
squared mass difference is $M_2^2$.  
For these values, the atmospheric $\nu_\mu$ deficit can be
explained by the $\psi_2\leftrightarrow \psi_3$ oscillation.
In the $(\psi_1, \psi_2)$ sector, the squared mass difference  
is $M_1^2$, which may realize the
appropriate values (1) and (2) for the MSW solution\cite{msw} of solar 
neutrino problem.
As to the mixing angle, both of the small mixing and large mixing
solution are possible depending on the ratio of $\mu_1$ and $\mu_2$.
Moreover, it is interesting that in the present scenario only nonzero 
mass eigenvalue $M\mu_4^2$ in the first approximation can be set just 
in the appropriate region for the 
hot dark matter, which can explain the structure formation of the 
universe.
The mass matrix $M_{\rm light}+M_{\rm per}$ is found to satisfy all 
preferable features for the neutrino problems.

It will be useful here to present
some comments related to the observation. 
In this scheme the solar neutrino deficit is explained by
$\nu_e \rightarrow \nu_s$\cite{es} and
the atmospheric neutrino deficit is due to the
$\nu_\mu$-$\nu_\tau$ oscillation.
The possibility of $\nu_e \rightarrow \nu_s$ will be examined by 
the future solar neutrino experiments in
Super-Kamiokande and SNO as pointed out in \cite{nucl1}.
Also the observation of $\nu_\mu$-$\nu_\tau$ oscillation 
will be helpful to distinguish this scenario from other possibilities.

It will be also interesting to consider in what kind of supersymmetric
models this type of mass matrix can be realized.
To see this point in more detailed, 
we concretely write down eq.(7)
in the case of $\mu_1 \ll \mu_2 \simeq \mu_3 \ll \mu_4$,
\begin{equation}
M_{\rm per}
\simeq
\left( \begin{array}{cccc}
A\mu_1^2 & A\mu_1\mu_2 & B\mu_1\mu_3 & -C\mu_1\mu_4 \\
A\mu_1\mu_2 & A\mu_2^2 & B\mu_2\mu_3 & -C\mu_2\mu_4 \\
B\mu_1\mu_3 & B\mu_2\mu_3 & A\mu_3^2 & -C\mu_3\mu_4 \\
-C\mu_1\mu_4 & -C\mu_2\mu_4 & -C\mu_3\mu_4 & 2C\mu_3^2 \\
\end{array}
\right),
\end{equation}
where
$$A\sim {1\over 2}(M_1+M_2)\mu_3^{-2}, \quad 
B\sim {1 \over 2}(-M_1+M_2)\mu_3^{-2}, \quad
C\sim M_2\mu_4^{-2}.$$
Here we should note that in eq.(13) there appears the analogous 
structure for $\mu_\alpha$ to eq.(5).
In eq.(5) $\mu_\alpha$ stands for an effective mixing
between $\psi_\alpha$ and $N_R$.
If $M_{\rm per}$ is induced as the loop effects through the
suitable Yukawa couplings,
the factors $\mu_\alpha\mu_\beta$ in eq.(13) can be interpreted 
just as the ratio of the product of corresponding Yukawa couplings 
which compose the loop diagrams. 
This feature at least suggests that compared with the MSSM contents, the 
necessity of the extension of doublet Higgs sector and also the
introduction of the singlet sector 
which has the couplings with other fields similar to the ones of a 
heavy right-handed neutrino $N_R$.\footnote{The trial in this
direction is presented in ref.\cite{s}.}
Related to the model construction,
we can settle the parameters in eq.(4) to
satisfy eqs.(11) and (12) in the case of $M\sim 10^{12+2q}$~GeV,
\begin{eqnarray}
{\rm (i)}&&m_4\sim 10^{2+q},\quad  m_3\sim m_2\sim 10^{2+q-p},\quad  
\quad m_1\sim 5\times 10^{q-p}, \nonumber \\
{\rm (ii)}&&m_4\sim 10^{2+q}, \quad m_3\sim m_2\sim m_1\sim 
10^{2+q-p}, 
\end{eqnarray}
where these values are written in the GeV unit.
A constant $q$ is a free parameter here but in principle it will be
determined by fixing the generating mechanism of a heavy right-handed
neutrino mass $M$. For the concrete model construction
it may be helpful to note the following points related to the parameter
$q$. 
$M\mu_2^2\sim M\mu_3^2\sim 10^{1-2p}$eV is always satisfied
independently of the value of $q$
and also the value $m_4\sim 1$~GeV corresponds to  $M\sim 10^8$~GeV($q=-2$).

As is easily seen from the formulus (6), 
under the above parameter settings (11) or (14), 
the mixings between $(\psi_1, \psi_2, \psi_3)$ and $\psi_4$ are 
crucially dependent on the parameter $p$.
The osillations between them are very important from various
viewpoints. 
Relating to this point, we should add here some discussions about 
the probability that the
sterile neutrino becomes the dark matter candidate without conflicting
with the BBN bound\cite{nucl1,dw}.
We consider the neutrinos $\nu_L$ and $\nu_R$ whose mass terms are 
given by
\begin{equation}
-{\cal L}_{\rm mass}=m_D\bar\nu_L\nu_R+{1 \over 2}m_R\nu_R\nu_R +h.c.,
\end{equation}
where we assume $m_D \ll m_R$ and then the light neutrino mass is given 
as $m_\nu=m_D^2/m_R$.
Following the analysis in ref.\cite{dw}, the relation between the distribution 
functions $f_s, f_\nu$ of sterile and active neutrinos at the period of
nucleosynthesis is given by
\begin{equation}
{f_s \over f_\nu}={6.0\over g^{1\over 2}_*}\left({m_D \over 1~{\rm eV}}
\right)^2\left({1~{\rm keV} \over m_R}\right),
\end{equation} 
where $g_*$ is the number of effective massless degrees 
of freedom at that time.
In order to estimate $m_D$ and $m_R$ consistent with the BBN we
take account of the relations
\begin{equation}
{\Omega_s \over \Omega_\nu}={m_Rf_s \over m_\nu f_\nu}, \qquad 
{m_\nu \over \Omega_\nu}\simeq 92h^2~{\rm eV}
\end{equation}
where $h=H_0/(100{\rm km}~{\rm sec}^{-1}~{\rm MPc}^{-1})$ and $H_0$ is
Hubble constant today.
The ratio of energy density of particle species $a$ to the critical
energy density of the universe is represented as 
$\Omega_a \equiv \rho_a/\rho_c$. From eqs.(16) and (17) we get
\begin{equation}
m_D \sim 1.2\times 10^{-1}g_*^{1\over 4}\Omega_s^{1\over 2}h.
\end{equation}
Following BBN scenario, the contribution of sterile neutrinos to the 
energy density at the time of primodial nucleosynthesis must be smaller 
than the contribution of a light neutrino species : 
$f_s ~{^<_\sim}~0.4f_\nu$.
From this fact, using (17) we get
\begin{equation}
m_R~{^>_\sim}~230h^2\Omega_s~{\rm eV}.
\end{equation}
In case of $m_R\gg m_D$, the mixing angle $\theta$ is given by
$\theta\sim m_D/m_R$ and then
\begin{equation}
m_R^2\sin^2\theta \sim m_D^2\sim 1.5\times 10^{-2}h^2g_*^{1\over 2}\Omega_s~
{\rm eV}^2.
\end{equation}
If we take $g_*\sim 10.8$, $h\sim 0.5$ and $\Omega_s\sim 0.3$ as the typical
values, we can roughly estimate $m_R$ and $m_R^2\sin^2\theta$ as 
$m_R~{^>_\sim}~17$~eV and $m_R^2\sin^2\theta\sim 10^{-2}~{\rm eV}^2$.

This result can be applyed to $(\psi_3, \psi_4)$ sector of the present 
model when the sterile neutrino $\psi_4$ is regarded as the dark 
matter candidate.\footnote{In the present model $(\psi_2, \psi_4)$ sector 
has also the similar mixing to $(\psi_3, \psi_4)$ sector. If we 
consider its effect, the mass bound for the sterile neutrino will
become larger.}
In fact, focussing on $(\psi_3,\psi_4)$ sector in $M_{\rm
light}+M_{\rm per}$ and noticing that 
$M\mu_4^2$ and $M\mu_3\mu_4$ can be 
corresponded to $m_R$ and $m_R\sin\theta$ in the above arguments,
we find some interesting features on the basis of the above BBN constraints.
At first, if we take $p \sim 2$ under such constraints, 
we can set a value of $M_2$ freely in the
range which is appropriate
for the explanation of atmospheric neutrino problem independently of
the mass of the dark matter candidate $\psi_4$.
This is because the present mass perturbation $M_{\rm per}$ 
is such one that does not disturb the mixing property (6).
On the other hand, from eq.(19) the sterile neutrino mass should be 
${^>_\sim}$~17~eV as suggested in \cite{nucl1,dw},
which is somehow larger value than one of the usual hot dark 
matter\cite{dark,phkc}.\footnote{
If $\nu_{\tau}$ is considered to be a dark matter candidate, 
such a constraint will not appear.}

Some remarks related to this scenario are ordered.
Firstly we should comment on the possibility to embed the present scenario 
into some underlying theories.
We have already presented some statements on this point.
Such models will not be the usual grand unified models with the ordinary
seesaw mechanism, in which the right handed neutrinos are required to be 
all heavy.  
One promising possibility is a superstring inspired $E_6$ model, in which the 
group theoretical constraints on the Yukawa couplings become very weak. 
Usually it is not so easy to accomodate small neutrino masses and induce 
neutrino oscillation without bringing other phenomenological
difficulty in that framework\cite{esix}.
However, if we introduce unconventional field assignments\cite{nar} 
under suitable conditions in that model, it is possible to show that 
the similar mass matrix structure discussed here can be realized.
The related study of this subject will be presented elsewhere\cite{s}.

Next it may be useful to present a brief discussion on other possibility 
of the state identification such as 
$\psi_2 \rightarrow$ a right-handed sterile 
neutrino($\nu_s$) and
$\psi_1, \psi_3, \psi_4 \rightarrow \nu_{eL}, \nu_{\mu L}, \nu_{\tau
L}$.
In our scheme the light right-handed neutrino has the mixing with 
left-handed active neutrinos. 
The BBN again severely constrains the mixing angle $\theta$ and the 
squared mass difference $\Delta \tilde m^2$ 
for a sterile neutrino mixing with left-handed active 
neutrinos $\nu$\cite{nucl1,bd},
\begin{eqnarray}
&&\Delta \tilde m^2\sin^42\theta~{^<_\sim}~5\times 10^{-6}~{\rm
eV}^2~~(\nu=\nu_e),
\nonumber \\
&&\Delta \tilde m^2\sin^42\theta~{^<_\sim}~3\times 10^{-6}~{\rm eV}^2
~~(\nu=\nu_{\mu ,\tau}).
\end{eqnarray}
These constraints rule out the large mixing MSW solution of solar neutrino
problem due to $\nu_e\rightarrow\nu_s$ and also the explanation of
atmospheric neutrino problem by $\nu_\mu\rightarrow\nu_s$.
These constraints are derived under the assumption that
the relic neutrino asymmetry $L_\nu$ is very small.
Although it has been recently suggested in ref.\cite{nucl2} that 
for $L_\nu >7\times 10^{-5}$ both
of these solutions can be consistent with BBN constraints,
this seems to be unlikely realized.
The scenario studied here seems to be an only allowed one following the 
consistency with the BBN.

We consider here only a diagonal charged lepton mass matrix.
For the non-diagonal one like the Fritzsch mass matrix\cite{frit} 
the situation becomes more complicated and the
introduction of complex phases will be necessary 
to make our scenario available as suggested in ref.\cite{matrix}.
 
In summary we proposed a neutrino oscillation scenario based on a
suitable neutrino mass martix which can
explain successfully both of the deficits of solar neutrino and 
atmospheric neutrino within the framework of three generation left and right
handed neutrinos. Although in this scenario the suitable
mixings and only one nonzero mass eigenvalue are realized in the first 
approximation level, the appropriate effective 
squared mass differences
for the explanation of both problems can be induced through the mass
perturbation which does not violate the first approximated state mixings 
induced through the seesaw mechanism.
An interesting feature of this scenario is that it can naturally contain
a neutrino as the dark matter candidate with the appropriate mass.\\

The author is grateful for the hospitality of TH-Division of CERN
where the part of this work was done. 
He also thanks Dr.~C.~Giunti for his critical comments on 
the first version of the present paper. 
This work is supported
by a Grant-in-Aid for Scientific Research from the Ministry of Education, 
Science and Culture(\#05640337 and \#08640362).
 
\newpage
%%%%%%%%%%%%%%%%%%%%%%%%%%%% References %%%%%%%%%%%%%%%%%%%%%%%%%%%%%%%%%%%

\end{document}